\documentclass[letterpaper]{article}
\usepackage{iccc}
\usepackage{csquotes}
\usepackage{graphicx}
\usepackage{times}
\usepackage{helvet}
\usepackage{courier}
\title{Casual Source Code Editing}
\author{Ender Minyard\\
cum2102@columbia.edu\\
}
\begin{document} 
\maketitle
\begin{abstract}
\begin{quote}
  There has been substantial research undertaken on the role of computational systems that encourage autotelic creativity.
  Previous studies on the role of software in autotelic creativity have not explored code editing tools in much detail.  
  This study sets out to examine the role of code editing tools in autotelic creativity. 
  The principal findings of this research are that existing code editors can be adapted to support casual creativity, and that casual creators exhibit standard interaction design patterns. 
\end{quote}
\end{abstract}
\section{Introduction}
Studies on casual creators represent a growing field. 
Casual creators are systems that prioritize user experience over productivity while helping users of the system with creative tasks \cite{Compton2019CasualCD}.
Casual, or autotelic, creativity here is defined as \enquote{little-c} or \enquote{everyday innovation} creativity \cite{Kaufman2009BeyondBA}. 
Casual creativity produces artifacts that are unexceptional in most contexts but novel and useful to the user. \\ \\
Examples of design patterns featured within casual creators include \textbf{instant feedback} and \textbf{entertaining evaluations} \cite{Compton2019CasualCD}. 
The \textbf{instant feedback} pattern allows a user to see changes in an artifact immediately after manipulating casual creator controls. The \textbf{entertaining evaluation} pattern gives users feedback on artifacts co-created by users and the casual creator's generator. Attempts have been made to include casual creator design patterns in 
poetry generators \cite{Boggia2022CasualPC}, game designers \cite{Kreminski2020GerminateAM}, and knitting tools \cite{Graves2020eLoominateTF}. \\ \\
Source code describes how computer programs run. Source code editing is the process of editing computer programs.
Previous studies of casual creativity have not applied the casual creator framework to source code editing.
This study set out to apply the casual creator framework to the source code editing context.
It is hoped that this work will generate fresh insights into code editor design.
\section{Related Work}
A number of studies have begun to examine the role computing devices play in code editor design. Traditional code editors are designed for use with a keyboard, but this is an issue for mobile devices \cite{Raab2016SourceCI}.
Mobile devices are often the sole computing devices in countries where personal computers are less common \cite{Sukumar2019MobileDI} and generally dominate Internet traffic in countries where personal computers are more common \cite{krum_2022}.
These studies demonstrate the need for a code editor which runs on mobile devices.

To date, several studies have explored the relationship between device usage and Internet traffic. 
Mobile and desktop users search for similar content, but with different motivations. The strongest motivators for mobile web usage are awareness, time management, curiosity, diversion, social connection, and social avoidance \cite{Church2011UnderstandingMW}.
Increasingly, mobile web usage is intended to distract, but not to achieve any particular task \cite{Reis2012RethinkingMS}. Unlike personal computers, mobile devices are essentially social \cite{Church2011UnderstandingMW}.
These studies illustrate the importance of designing a code editor for mobile devices that provides distraction.

The problem of code editor design can be described as creating a user experience where coding is \enquote{an autotelic experience in itself} \cite{Compton2021ConversationSI}. This problem formulation aligns with the need for code editors that meet the mobile user's motivation to satisfy their curiosity and divert their attention.

Perhaps the most thorough account of casual code editing is to be found in the work of \cite{Compton2021ConversationSI}, who imagines a hypothetical code editor that \enquote{breaks free from productivity}. 
In this hypothetical code editor, \cite{Compton2021ConversationSI} describes how instant feedback can lead to users writing code extemporaneously. 
In \cite{McNutt2023ASO} instant feedback took the form of visualized program state. 
Visualizing program state could encourage creativity \cite{McNutt2023ASO}.

\section{System Description}
\begin{figure}
  \begin{center}
    \includegraphics[width=0.4\textwidth]{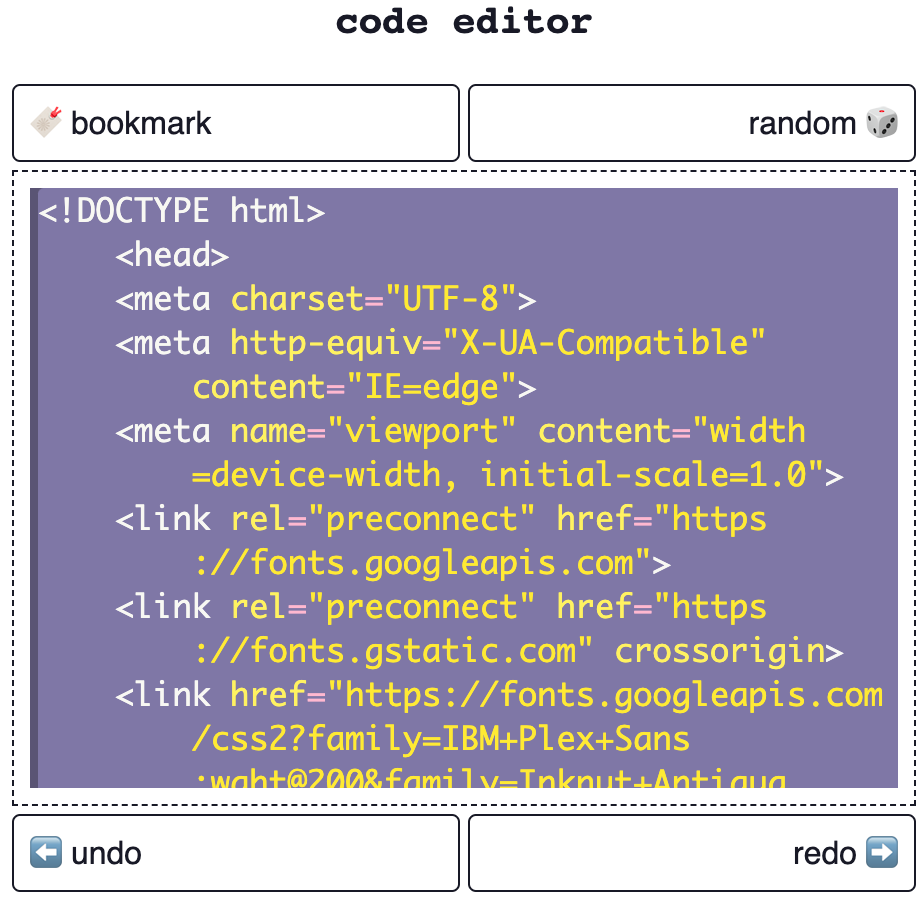}
    \caption{Code editing panel.}
    \label{fig:edit}
  \end{center}
  \end{figure}

\begin{figure}
  \begin{center}
  \includegraphics[width=0.4\textwidth]{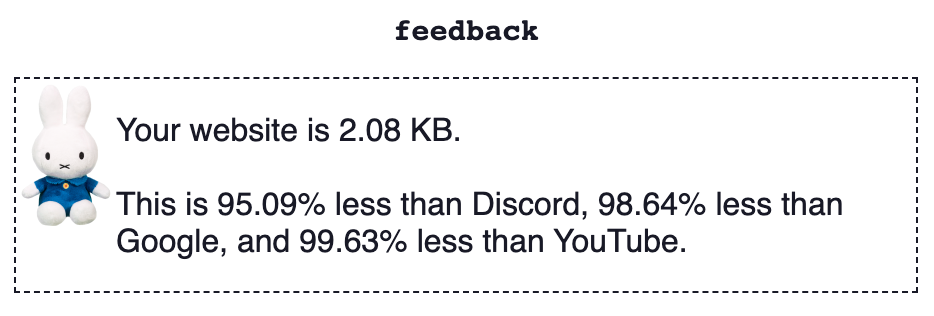}
  \caption{Feedback panel.}
  \label{fig:feedback}
\end{center}
\end{figure}

The system is a web application designed for casual source code editing in HyperText Markup Language, Cascading Style Sheets, and JavaScript, hosted on Github Pages. \footnote{https://github.com/enderminyard/phonecoding}
It is a full system, meaning it stands alone and is not part of a larger system.
The system is new. 
During its lifecycle, the system generates artifacts triggered by edits to the code editing panel (Figure \ref{fig:edit}) and user interaction with the \textbf{random} button (Figure \ref{fig:edit}). 
At any time, the user can generate code by pressing the \textbf{random} button at the code editing panel's upper right corner in Figure \ref{fig:edit}. \\
The system's goal is to enable autotelic coding. The system's generator is nondeterministic and tile-based \cite{Compton2021AGF}. One benefit of the system's simple tile-based generator is its low energy consumption on mobile devices. \\
The generator's slots consist of CSS properties and HTML elements. The generator's tiles consist of CSS properties to fill the CSS attribute slots, and text and images to fill the HTML element slots. The generator's selection logic is nondeterministic.
The generator's slots include the CSS properties for \verb|background-image|, \verb|border-radius|, \verb|font-style|, \verb|font-stretch|, \verb|font-style|, \verb|letter-spacing|, \verb|font-family|, \verb|filter|, \verb|translate|, and \verb|border|. The values for the tiles to fill these slots in the generator were chosen by using values displayed in the MDN Web Docs page for each CSS property. 
The HTML element slots in the generator consist of a \verb|span| element, an \verb|img| element, an \verb|alt| tag for the \verb|img| element, and a \verb|figcaption| element. The values for the tiles to fill these slots in the generator were chosen to be novel or potentially novel.  
When a user of the system presses the \textbf{random} button as shown in Figure \ref{fig:edit}, the generator is triggered to generate HTML. This HTML is displayed in both the code editing panel and the preview panel. \\
\begin{figure}
  \begin{center}
  \includegraphics[width=0.4\textwidth]{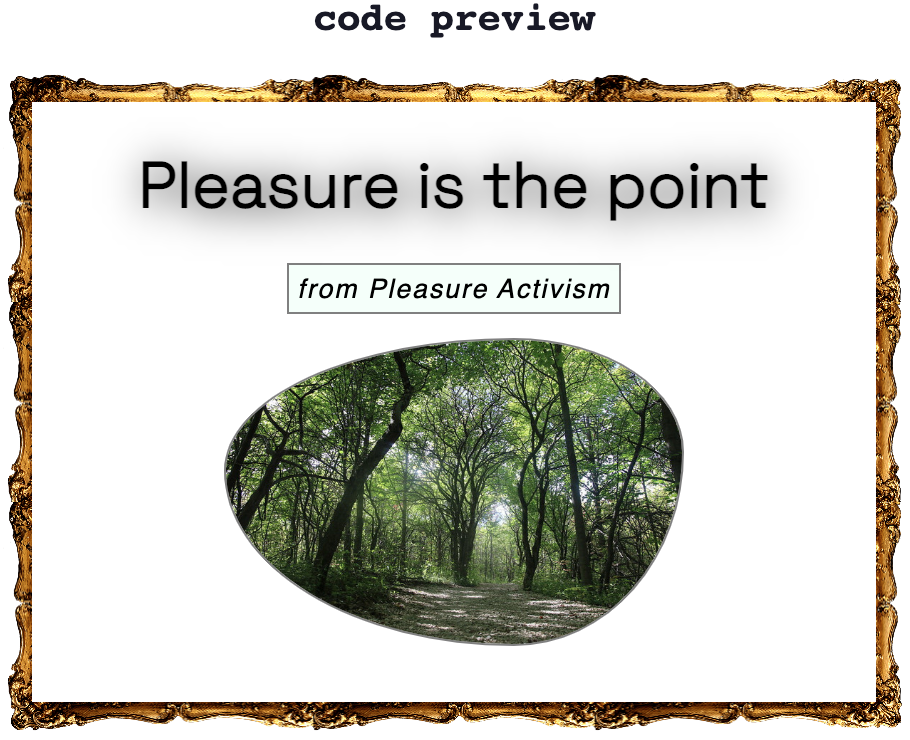}
  \caption{Code preview panel.}
  \label{fig:preview}
  \end{center}
\end{figure}
The system communicates with others when source code in the code editing panel in Figure \ref{fig:edit} is changed or when 
the user presses the \textbf{random}, \textbf{undo}, or \textbf{redo} buttons. These user interactions lead to a change in the system's display of state. 
The system is a code editor that displays system state in the code editing panel (Figure \ref{fig:edit}), code preview panel (Figure \ref{fig:preview}) and feedback panel (Figure \ref{fig:feedback}). \\

\section{Evaluation}

\subsection{Design Patterns}
Previous studies such as \cite{Kreminski2020GerminateAM} perform principles-based evaluation on the casual creator design patterns first proposed by \cite{Compton2015CasualC} and described in further detail within \cite{Compton2019CasualCD}.
\subsubsection{Instant feedback}
Users of the code editor can see changes to the system's generated artifact, HTML source code, in near real-time. All changes to 
the system's generated HTML source code are displayed in the code editing panel (Figure \ref{fig:edit}), code preview panel (Figure \ref{fig:preview}) and feedback panel (Figure \ref{fig:feedback}).
\subsubsection{Chorus Line}
The code editor does not allow a user to see multiple artifacts generated at once and compare the qualities present in each artifact. 
\subsubsection{Simulation and approximating feedback}
The code editor increases user confidence and pride in artifacts created using the system within a feedback panel that gives users partial feedback. The feedback panel informs users of the size of the generated HTML webpage (Figure \ref{fig:feedback}).
\subsubsection{Entertaining evaluations}
The code editor improves the user experience by giving users feedback regarding generated artifacts in a method that is novel.
This design pattern is implemented by comparing the size of webpages generated by the code editor to the size of webpages such as the Google homepage (Figure \ref{fig:feedback}).
\subsubsection{No blank canvas}
The code editor prevents users from being initimidated by a blank canvas by presenting new users with an artifact to use.
Generating a random HTML webpage on page load is an example of this design pattern.
\subsubsection{Limiting actions to encourage exploration}
Limiting the range of possible code edits encourages users of the code editor to feel safe exploring the possibility space of the system.
Traditional code editors allow users to add multiple HTML pages to one project, but this study's code editor limits users to one HTML page per project. 
\subsubsection{Mutant shopping}
\begin{figure}[h]
  \begin{center}
  \includegraphics[width=0.4\textwidth]{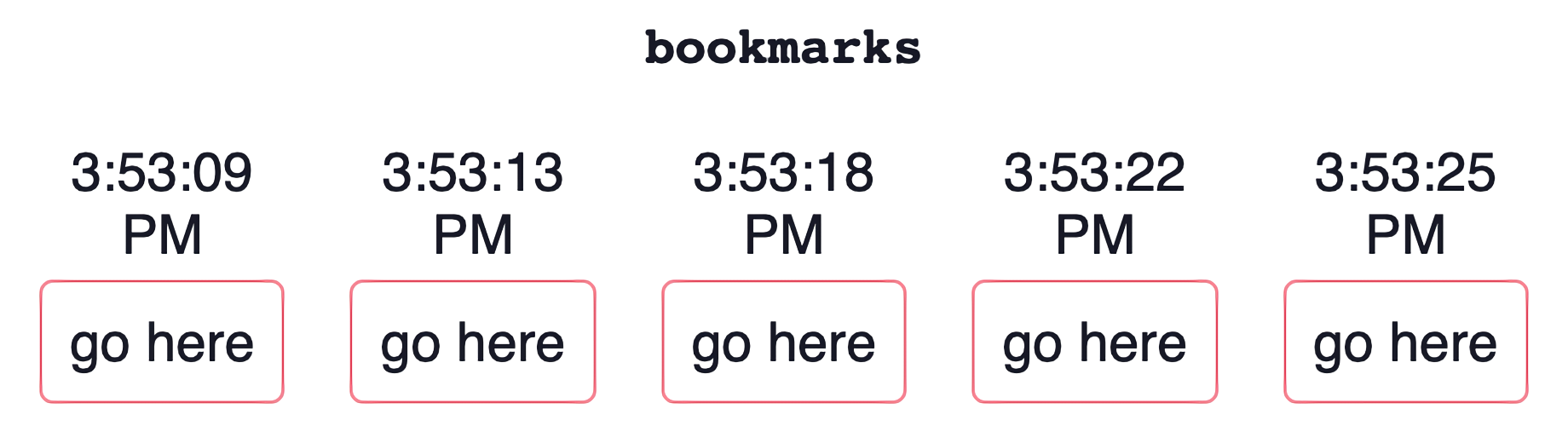}
  \caption{Bookmarking artifacts.}
  \label{fig:bookmark}
\end{center}
\end{figure}

The code editor allows users to browse the possibility space of the system's generator with a \textbf{random} button that generates HTML code when pressed (Figure \ref{fig:edit}). As shown in Figure \ref{fig:bookmark}, users can also bookmark artifacts for later examination.
\subsubsection{Modifying the meaningful}
The code editor does not restrict user modification of artifacts to actions that affect the resulting artifact using methods significant to the user.
\subsubsection{Saving and sharing}
\begin{figure}[h]
  \begin{center}
  \includegraphics[width=0.4\textwidth]{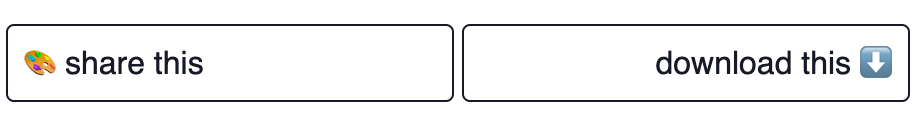}
  \caption{Saving and sharing.}
  \end{center}
  \label{fig:save}
\end{figure}

Saving and sharing artifacts generated by the system enables creativity to continue outside the boundaries of the code editor.
The code editor implements this design pattern with \textbf{share} and \textbf{save} buttons (Figure \ref{fig:save}).

\subsubsection{Hosted communities}
The code editor does not host an online community.
\subsubsection{Modding, hacking, teaching}
The code editor does not allow system users to modify the casual creator itself.
\subsection{Results}
The code editor realizes seven out of eleven casual creator design patterns. 
In a code editor, however, limiting the space of user actions to meaningful modifications alone may not be technically feasible using existing programming languages. 
The three remaining casual creator design patterns are technically feasible to implement. 
In future investigations, it might be possible to realize a different system in which an autotelic code editor allows a user to see multiple generated artifacts at once, hosts an online cumminty, and allows modifications to its own source code. \\
Of the seven casual creator design patterns realized by the code editor, five are realized by displaying system state.
\begin{table}[ht]
\centering
  \begin{tabular}{c|c|c}
    \textbf{Component} & \textbf{Patterns} & \textbf{State} \\
    Editing & 2 & Yes \\
    Preview & 1 & Yes \\
    Feedback & 3 & Yes \\
    Random & 2 & No \\
    Bookmark & 1 & Yes \\
    Save & 1 & No \\
    Share & 1 & No
  \end{tabular}
  \caption{The code editor components displaying state implement the majority of the casual creator design patterns.}
\label{tab:state}
\end{table}

What can be clearly seen in Figure \ref{tab:state} is the positive correlation between state visibility and design pattern realization. 
What emerges from the results reported here is that visibility of system status \cite{Nielsen1994HeuristicE} may encourage casual creativity.
\section{Discussion}
One of the aims of this study was to bring the casual creator framework into the source code editing context. 
The results of this study indicate that casual creator design patterns encourage sharing system state with users.
The current study found that it is possible to edit code within a casual creator using existing programming languages. Existing code editors can be adapted to support casual creativity by visualizing program state, generating code samples on demand, and enabling code sharing.
This finding is contrary to \cite{Compton2019CasualCD} which has suggested that enabling code editing within a casual creator would require the adoption of new programming languages. \\ \\
% integrate casual creator patterns into source code editing
As mentioned in the literature review, mobile devices generally dominate Internet traffic \cite{krum_2022}. 
Mobile users are less task-oriented than desktop users \cite{Reis2012RethinkingMS}. 
The code editor presented in this study is designed to encourage the \textit{autotelic} completion of coding tasks on mobile devices.
An implication of this is the possibility that mobile users would use this code editor out of curiosity or boredom.
It can thus be suggested that the code editor presented within this study enables \textit{casual coding}.

\section{Conclusion}
This paper has argued that code editing could be brought to mobile devices in the form of a casual creator. The aim of the present research was to examine the viability of code editing in the casual creator context. 
In general, it seems that existing code editors can be adapted to encourage casual creativity.

Visibility of system status emerged as a reliable predictor for whether or not a system qualified as a casual creator.
This finding suggests that following standard interaction design patterns, such as showing the visibility of a system's status \cite{Nielsen1994HeuristicE}, may encourage casual creativity.

\bibliographystyle{iccc}
\bibliography{iccc}
\end{document}